\documentclass{ws-p10x7}

\begin{document}

\title{Thermal forward scattering amplitudes in temporal gauges}

\author{F. T. Brandt, J. Frenkel and F. R. Machado}
\address{Instituto de F\'\i sica,
Universidade de S\~ao Paulo\\
S\~ao Paulo, SP 05315-970, BRAZIL}

\twocolumn[\maketitle\abstract{
We employ the thermal forward scattering amplitudes technique in order to
compute the gluon self-energy in a class of temporal gauges. The
leading $T^2$ and the sub-leading $\ln(T)$ contributions are obtained 
for temperatures high compared with the external momentum. 
The logarithmic contributions have the same structure as the 
ultraviolet pole terms which occur at zero temperature (we have
recently extended this result to the Coulomb gauge).
We also show that the prescription poles, characteristic of temporal gauges,
do not modify the leading and sub-leading high-temperature behavior.
The one-loop calculation shows that the thermal self-energy
{\it is transverse}. This result has also been extended to higher
orders, using the BRS identities}]

There have been many investigations of thermal gauge field theories
in the temporal gauge, both in the imaginary and in the real time formalisms
\cite{kajantie:1985xx,kobes:1989up,james:1990it,leibbrandt:1994ki}. 
One of the main advantages of the 
{\it non-covariant} temporal gauge is that it is physical and
effectively ghost-free. At finite temperature, it may be considered
a more natural choice, since the Lorentz invariance is already
broken by the presence of the heat bath. It is also convenient
for calculating the response of the QCD plasma to a chromo-electric
field \cite{kajantie:1985xx,lebellac:book96}.
Despite these advantages, explicit calculations are known to be more
involved than in covariant gauges, mainly because of the extra poles
at $q\cdot n=0$ in the propagator, where $q$ is the loop momentum and
$n=(n_0,\vec 0)$ $(n_0^2 > 0)$ is the temporal axial four-vector. 

The standard method of calculation in the imaginary time formalism, 
employs the {\it contour integral formula} \cite{lebellac:book96}
\begin{eqnarray}\label{contour}
&T\displaystyle{\sum_{n=-\infty}^{\infty}}
f(q_{0}=i\,\omega_n^\sigma)=\nonumber\\
&\displaystyle{1\over 2\pi i}\displaystyle{\oint_{C}}dq_{0}f(q_{0}){1\over2}
\left[\coth\left({1\over2}\beta q_{0}\right)\right]^{i^{(2\sigma)}},
\end{eqnarray}
where
\[
{ \omega_n^\sigma=\pi\,T\,(2 n + \sigma)}; 
\left\{
\begin{array}{ll}
{\sigma=0} & (Bosons) \\
{\sigma=1} & (Fermions)
\end{array}
\right.
\]
and the contour $C$ is formed by two anti-parallel straight lines
equidistant from the imaginary axis. 
Of course, this formula cannot be employed when the function $f(q_0)$
has poles along the imaginary axis. This is the reason why one should
use some prescription for the poles in the temporal 
gauge \cite{leibbrandt:1994ki} gluon propagator
\begin{eqnarray}\label{propag}
\frac{1}{q^2}\left\{-i\delta^{ab}\left[g_{\mu\nu}-{1\over q\cdot u}
(q_{\mu}u_{\nu}+q_{\nu}u_{\mu}) + \right.\right. \nonumber \\
\left.\left. {q_{\mu}q_{\nu}\over (q\cdot u)^{2}}
\left({\alpha\over n_{0}^{2}} q^{2}+1\right)\right]\right\} \, ,
\end{eqnarray}
where $u=n/n_0$ is the heat bath four-velocity.

For general covariant gauges, it is possible to show that all the thermal Green
functions can be expressed (after the Cauchy integration in the complex
plane) in terms of forward scattering amplitudes of on-shell thermal
particles  \cite{frenkel:1990br1991ts,brandt:1997se}.
The main purpose of this work is to extend the forward scattering method to
a class of temporal gauges. As an illustration of this technique, we
will compute the {\it full tensor structure} of the one-loop gluon self-energy.
(Previous calculations have considered only the static limit of the
component $\Pi_{00}$ of the self-energy 
\cite{kajantie:1985xx,leibbrandt:1994ki}). In this way, 
we will be able to investigate the properties of transversality and
gauge invariance of the leading contributions proportional to
$T^2$. We also verify that {\it the gauge dependent} sub-leading logarithmic
part shares with the previous calculations in general covariant 
gauges \cite{brandt:1997se} the interesting property of having 
the same structure as the {\it ultraviolet pole contributions} which 
occur at zero temperature. 

The details involved in the forward scattering technique are
explained in the appendix A of reference \cite{brandt:2000a}. An
important condition in order to be able to apply this technique is
that the gluon propagator should have only the mass-shell poles at
$q^2=0$. Therefore, our first task when using the axial gauge gluon
propagator is to separate the contributions which can potentially have
poles at $n\cdot q=0$ from the normal mass-shell poles. The simplest
contribution having poles only at $n\cdot q=0$ is the ghost loop
diagram shown in Fig. 1(c).
\begin{figure*}
\epsfxsize30pc
\figurebox{10pc}{20pc}{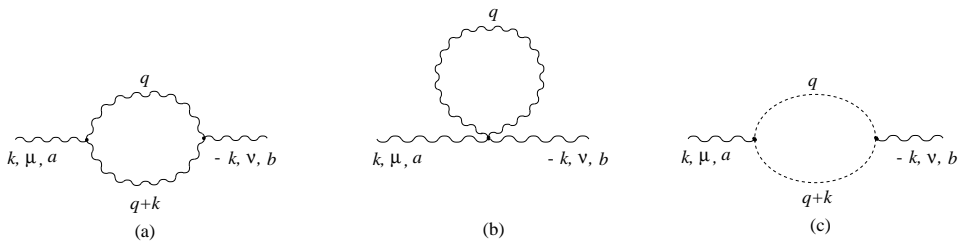}
\caption{One-loop diagrams which contribute to the gluon-self
energy. Wavy and dashed lines denotes respectively gluons and
ghosts. All external momenta are inwards.} \label{fig1}
  \end{figure*}
At finite temperature it is proportional to
\begin{equation}\label{ghost}
\int d^{3}\vec q \sum_{q0} \left[{t_{\mu\nu}\over 
n\cdot q\,n\cdot (q+k)}+q\leftrightarrow - q\right],
\end{equation}
where $t_{\mu\nu}$ is a momentum independent quantity and 
$q_0=2\pi\, i\, n\, T$ ($n=0,\pm 1, \pm 2,\cdots$).
Using partial fractions the integrand in (\ref{ghost}) can be wirtten as 
\begin{equation}\label{parfrac}
{1\over n\cdot k}\left[{1\over n\cdot q}-{1\over n\cdot (q+k)}\right].
\end{equation}
Performing a shift $q\rightarrow q-k$ in the second term,
we can easily see that the ghosts {\it effectively decouple}.
This shows that there is a simple mechanism for 
{\it some} cancellations of the temporal gauge poles, which involves
only simple algebraic manipulations, {\it before}
the computation of the sum over the Matsubara frequencies $q_0$.
As we will see next, there are other contributions
from the diagrams in  Figs. 1(a) and 1(b) sharing same property. 

The separation of the temporal gauge poles can be accomplished in a more
systematic and physical way, using the following well known tensor 
decomposition of the gluon self-energy
\begin{eqnarray}\label{eq7a}
\Pi^{ab}_{\mu\nu}=
\delta^{ab}\left(\Pi_{T} P_{\mu\nu}^T + \Pi_{L} P_{\mu\nu}^L
\right.\nonumber\\ \left.
+\Pi_{C} P_{\mu\nu}^C + \Pi_{D} P_{\mu\nu}^D\right),
\end{eqnarray}
where
\begin{eqnarray}\label{eq7}
P_{\mu\nu}^T & = & g_{\mu\nu}-P_{\mu\nu}^L-
P_{\mu\nu}^D \, ,
\nonumber \\
P_{\mu\nu}^L & = & \frac{\left(u\cdot k k_\mu -k^2 u_\mu\right)
                          \left(u\cdot k k_\nu -k^2 u_\nu\right)}
{-k^2|\vec k|^2}\, , 
\nonumber \\
P_{\mu\nu}^C & = & {2k\cdot uk_{\mu}k_{\nu}-k^{2}(k_{\mu}u_{\nu}+k_{\nu}u_{\mu}) 
\over k^{2}|\vec k|} \, , \nonumber 
\nonumber \\ 
P_{\mu\nu}^D & = & {k_{\mu}k_{\nu} \over k^{2}} \, ,
\end{eqnarray}
where $k^\mu P_{\mu\nu}^{T,L}=0$, $k^i P_{i\nu}^{T}=0$, 
$k^i P_{i\nu}^{L}\neq 0$ ($i=1,2,3$) and $k^\mu\,P_{\mu\nu}^{C,D}\neq 0$.

The calculation of structures 
$\Pi_{A}$, $\Pi_{B}$, $\Pi_{C}$ and $\Pi_{D}$ and the following algebraic
manipulations were performed with the help of computer algebra. 
After partial fraction decompositions 
[as in Eq. (\ref{parfrac})] and
shifts $q\rightarrow q-k$, the rather involved expressions for
$\Pi_{C}$ and $\Pi_{D}$ simplify considerably and all the temporal
gauge poles at $n\cdot q=0$ cancel. We then proceed using the contour
integral formula [Eq. (\ref{contour})] and show that $\Pi_{C}=\Pi_{D}=0$.
It is interesting to note that, from the general 
relation \cite{weldon:1996kb}
\begin{equation}\label{Teq8}
\Pi_{D}={\Pi_{C}^{2}\over k^{2}-\Pi_{L}}
\end{equation}
the vanishing of $\Pi_{C}$ to one loop order implies, in fact, 
that $\Pi_{D}$ should vanish up to the three-loops order. 
Using the Becchi-Rouet-Stora identities \cite{becchi:1974md} we have
extended this result to all orders \cite{brandt:2000a}.

For the structures $\Pi_A$ and $\Pi_B$ the temporal gauge 
poles do not cancel at the integrand level and we have to employ a 
prescription in order to be able to use the
Eq. (\ref{contour}). Using the procedure described in the appendix of
reference \cite{leibbrandt:1994ki} we were able to show, by explicit
calculation, that the {\it prescription poles} do not contribute to
the leading and the sub-leading high temperature limit of $\Pi_A$ and
$\Pi_B$. Therefore, the thermal gluon 
self-energy can be represented, in the limit of high temperatures, 
in terms of  {\it forward scattering amplitudes} of on-shell thermal 
gluons, as given by Eq. (6) of reference  \cite{brandt:2000a}.

In conclusion, our results show that the full tensor structure of the
thermal gluon self-energy can be consistently computed in a class of
temporal gauges, and expressed in terms of forward
scattering amplitudes of on-shell thermal gluons. Using this approach,
we have obtained the known gauge invariant result for the leading
high temperature $T^2$ contribution. We also have shown, by explicit
calculation, that the one-loop thermal self-energy 
{\it is exactly transverse} for any temperature regime. 
Motivated by this result, we were able to prove the 
transversality to all orders.
This property seems to be very peculiar to the temporal
gauges. It is not valid, for instance, in general covariant gauges, 
except for the Feynman gauge, where the transversality has been 
verified only to one-loop order \cite{brandt:1997se}.

Our approach also gives sub-leading contributions
which are in agreement with the conjecture proposed in
\cite{brandt95}, according to which the ultraviolet
divergent contributions  which arises at $T=0$ are identical to the
thermal contributions proportional to $\ln(1/T)$.
We have recently verified this conjecture also in the {\it Coulomb gauge}
so that the divergent part of the Coulomb gauge gluon 
self-energy \cite{ft76,lw96} can be alternatively
obtained from our $\ln(1/T)$ contribution. The details of this
analysis will be reported elsewhere.

\section*{Acknowledgments}
F.T.B and F.R.M. acknowledge the financial support from FAPESP (Brazil).
F.T.B and J.F. would like to thank CNPq for a grant.


\begin{thebibliography}{99}
\bibitem{kajantie:1985xx}
K. Kajantie and J. Kapusta, Ann. Phys. {\bf 160},  477  (1985).
\bibitem{kobes:1989up}
R. Kobes, G. Kunstatter, and K.~W. Mak, Z. Phys. {\bf C45},  129  (1989).
\bibitem{james:1990it}
K.~A. James and P.~V. Landshoff, Phys. Lett. {\bf B251},  167  (1990).
\bibitem{leibbrandt:1994ki}
G. Leibbrandt and M. Staley, Nucl. Phys. {\bf B428},  469  (1994).
\bibitem{lebellac:book96}
M.~L. Bellac, {\em Thermal Field Theory} (Cambridge University Press,
  Cambridge, England, 1996).
\bibitem{frenkel:1990br1991ts}
J. Frenkel and J.~C. Taylor, Nucl. Phys. {\bf B334},  199 (1990);
{\bf B374},  156  (1992).
\bibitem{brandt:1997se}
F.~T. Brandt and J. Frenkel, Phys. Rev. D {\bf 56},  2453  (1997).
\bibitem{brandt:2000a}
F.~T. Brandt, J. Frenkel and F.~R. Machado, 
Phys. Rev. D {\bf 61}, 125014 (2000).
\bibitem{weldon:1996kb}
H.~A. Weldon, Annals Phys. {\bf 271},  141  (1999).
\bibitem{becchi:1974md}
C. Becchi, A. Rouet, and R. Stora, Commun. Math. Phys. {\bf 42},  127
(1975).
\bibitem{brandt95}
F.~T. Brandt and J. Frenkel, Phys. Rev. Lett. {\bf 74}, 1705 (1995).
\bibitem{ft76}
J. Frenkel, J.C. Taylor
Nucl. Phys. {\bf B109}, 439 (1976).
\bibitem{lw96}
G. Leibbrandt and J. Williams,
Nucl. Phys. {\bf B475}, 469 (1996).

\end{thebibliography}
\end{document}